\documentstyle[preprint,psfig,aps]{revtex}
\begin{document}
\draft
\preprint{\begin{tabular}{l}
\hbox to\hsize{ \hfill BROWN-HET-1182}\\[-3mm] 
\hbox to\hsize{ \hfill KIAS-P99026}\\[5mm] \end{tabular}}
\bigskip

\title
{Higgs Boson Bounds in Three and Four Generation Scenarios}
\author{David Dooling$^{a}$, Kyungsik Kang$^{a}$ and Sin Kyu Kang$^{b}$}

\address{a. Department of Physics\\
Brown University, Providence RI 02912, USA \\
b. School of Physics \\
Korea Institute for Advanced Study \\
Seoul 130-012, Korea}
\maketitle

\tightenlines
\begin{abstract}
In light of recent experimental results, we present updated bounds on the lightest Higgs boson mass in the Standard Model (SM) and in the Minimal Supersymmetric extension of the Standard Model (MSSM). The vacuum stability lower bound on the pure SM Higgs boson mass when the SM is taken to be valid up to the Planck scale lies above the MSSM lightest Higgs boson mass upper bound for a large amount of SUSY parameter space.
If the lightest Higgs boson is detected with a mass $M_{H} < 134$ GeV (150 GeV) for a top quark mass $M_{top}$ = 172 GeV (179 GeV), it may indicate the existence of a fourth generation of fermions.
The region of inconsistency is removed and the MSSM is salvagable for such values of $M_{H}$ if one postulates the existence of a fourth generation of leptons and quarks with isodoublet degenerate masses $M_{L}$ and $M_{Q}$ such that 60 GeV $< M_{L} <$ 110 GeV and  $ M_{Q}>$ 170 GeV.
\end{abstract}

\bigskip
\centerline {\rm\bf I. Introduction}
 The search for the Higgs boson being one of the major tasks along with that for supersymmetric sparticle and fourth generation fermions at future accelerators such as LEP200 and LHC makes it a theoretical priority to examine the bounds on the Higgs boson mass in the SM and its supersymmetric extension and to look for any distinctive features.
 The actual measurement of the Higgs boson mass could serve to exclude or at least to distinguish between the SM(3,4) and the MSSM(3,4) models for electroweak symmetry breaking.
Recently, bounds on the lightest Higgs boson mass were calculated in \cite{1,2,3,4,5,6,7,8,9}.
It was found that for a measured $M_{H}$ lying in a certain mass range, both the SM vacuum stability lower bound and the MSSM upper bound are violated, thus shaking our confidence in these theories just as the final member of the mass spectrum is observed.
One method of curing this apparent illness is to take a leap of faith by adding another fermion generation, to fortify these theories with another representation of the gauge group.
This additional matter content, for certain ranges of its mass values, has the desired effect of raising the MSSM3 upper bound above that of the SM lower bound and avoids the necessity of being forced to introduce completely new physics.

Since our previous work \cite{1}, a new experimental lower bound on $M_{B}$, the fourth generation bottom-type quark, has become available from the CDF collaboration.
The new lower bound on $M_{B}$ of $\sim$ 140 GeV necessitates a reevaluation of the analysis in Ref. 1, in which we considered a completely degenerate fourth generation with mass $M_{4}$ and obtained an upper bound on $M_{4}$ of $\sim$ 110 GeV from considerations of gauge coupling unification.
In this work, we shall consider a fourth generation of fermions where degeneracy holds among the lepton and quark isodoublets separately.
This lifting of the complete degeneracy in the present analysis will enable us to make a definite and stronger statement about the allowed ranges of $M_{L}$.
In addition, we use the most recent experimental values for $M_{Z}$ and $\alpha(M_{Z})$.
Our presentation is organized as follows.
Bounds on $M_{H}$ are obtained by imposing different boundary conditions on the Higgs self-coupling $\lambda$.
We present the results with three generations of fermions and then we see how the potential measurements of $M_{H}$ that lead to mutual inconsistencies in the SM and in the MSSM can be accomodated naturally in the MSSM4. 
Our analysis yields a relatively narrow allowed range for $M_{L}$, the mass of the fourth generation leptons, that is consistent with gauge coupling unification.

The method of solving the RGE's and the appropriate boundary conditions for the couplings is explained in Ref. 1.
In this update, we use the same notation and procedure found in Ref. 1.
We also use the following values for $M_{Z}$ and $\alpha_3({M_{Z}})$: $M_{Z} = 91.1867$ GeV and $\alpha_{3}(M_{Z}) = .119$.

\section{Bounds on $M_{H}$}

We now determine a lower bound on the Higgs boson mass in the SM \cite{5,12}.
We first alert the reader to our phenomenologically viable assumption that the physical vacuum corresponds to a global, not merely a local, minimum of the effective potential.
This assumption is consistent with our intention to accept the SM as a truly valid theory and compute the consequences, i.e. to zeroth order there is no motivation to consider the physical vacuum to be anything other than the true vacuum.
If one considers the possibility that the physical vacuum is a metastable vacuum with a lifetime longer than the age of the universe, that there exist deeper minima of the potential, then the SM lower bounds on the Higgs boson mass become less stringent in general for certain choices of $\Lambda$ and $M_{top}$, where $\Lambda$ is the cutoff beyond which the SM is no longer valid \cite{13}.
But, for $M_{top} \sim 177$ GeV and $\Lambda = 10^{19}$ GeV, the SM3 absolute stability lower bound is relaxed by only $\sim O(5)$ GeV when one only imposes metastability requirements, and this small effect only becomes diminished with the inclusion of a fourth generation.
We obtain lower limits on the SM Higgs boson mass by requiring stability of this observed vacuum.
It is well known that lower values of $\Lambda$ relax the SM lower bounds \cite{15}, but we note that the lower bounds on the SM Higgs boson mass are insensitive to the precise value of $\Lambda$ for large $\Lambda$, i.e. for $10^{11}$ GeV $< \Lambda < 10^{19}$ GeV.
 
Working with the two-loop RGE requires the imposition of one-loop boundary conditions on the running parameters \cite{10}.
As pointed out by Casas et al. \cite{5,7}, the necessary condition for vacuum stability is derived from requiring that the effective coupling $\tilde{\lambda}(\mu)>$ 0  rather than $\lambda > 0$ for $\mu(t) < \Lambda$, where $\Lambda$ is the cut-off beyond which the SM is no longer valid. 
The effective coupling $\tilde{\lambda}$ in the SM4 is defined as:
\begin{displaymath}
\tilde{\lambda}=\frac{\lambda}{3} -\frac{1}{16 \pi^{2}}\left\{ \sum_{i=1}^{5} 2 \kappa_{i} h_{i}^{4} \left[ \ln \frac{h_{i}^{2}}{2} - 1 \right] \right\}
\end{displaymath}
where the three generation case is simply the same as the above expression without the fourth generation Yukawa coupling contributions.
Choosing $\Lambda = 10^{19}$ GeV and $M_{top} = 172$ GeV, we arrive at a vacuum stability lower bound on $M_{h}$ of $\sim$ 134 GeV for the SM with three generations.
Allowing $M_{top}$ to be as large as 179 GeV increases the lower bound on $M_{H}$ to $\sim$ 150 GeV.

To compute the MSSM upper bound on $M_{H}$, we assume that all of the sparticles have masses $O(M_{susy})$ or greater and that of the two Higgs isodoublets of the MSSM, one linear combination is massive, also with a mass of $O(M_{susy})$ or greater, while the other linear combination, orthogonal to the first, has a mass of the order of weak-scale symmetry breaking.
With these two assumptions, it is clear that below the supersymmetry breaking scale $M_{susy}$, the effective theory is the SM.
This fact enables us to use the SM effective potential for the Higgs boson when we treat the lightest Higgs boson in the MSSM.

In the MSSM(3,4), the boundary condition for $\lambda$ at $M_{susy}$ is
\begin{displaymath}
\frac{\lambda}{3}(M_{susy})=\frac{1}{4}\left[g_{1}^{2}(M_{susy})+g_{2}^{2}(M_{susy})\right]\cos^{2}(2\beta)+\frac{\kappa_{i}h_{i}^{4}(M_{susy})}{16 \pi^{2}}\left( 2\frac{X_{i}}{M_{susy}^{2}}-\frac{X_{i}^{4}}{6 M_{susy}^{4}} \right)
\end{displaymath}
where $\kappa_{i}$ = 3 for $i = (t,T,B)$ and $\kappa_{i}$ = 1 for $i = (N,E)$ and $X_{i}$ is the supersymmetric mixing parameter for the ith fermion.
Zero threshold corrections correspond to $X_{i}$ = 0.
Maximum threshold corrections occur for $X_{i} = 6 M_{susy}^{2}$.
\bigskip
\bigskip
\vglue -1cm
\hglue -2cm
\psfig{figure=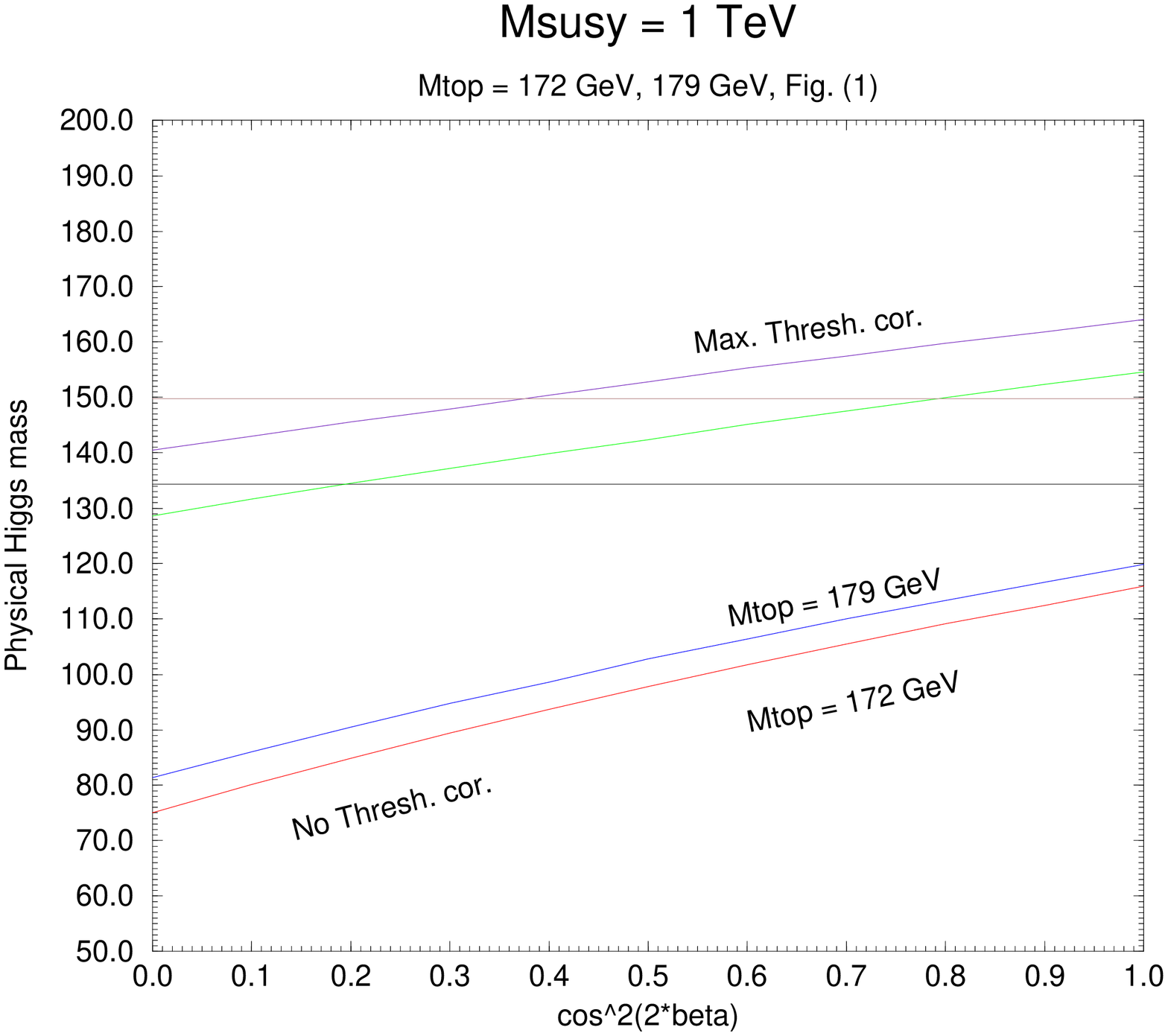,height=8cm,angle=-90}\hglue 1cm
\psfig{figure=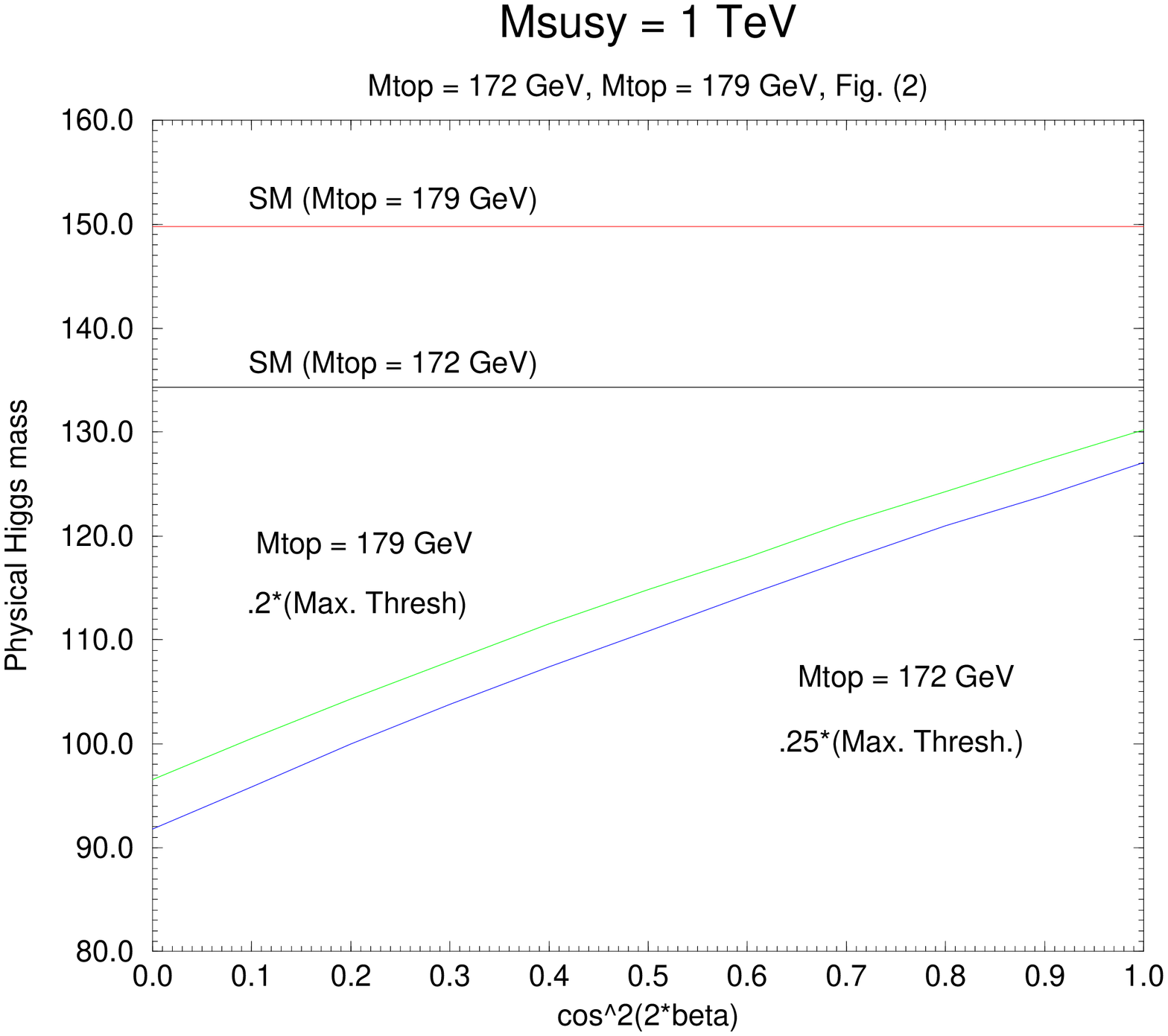,height=8cm,angle=-90}
\hglue 3.5cm (1)\hglue 10cm (2)\\
\begin{quote}
\scriptsize Figure 1: The lightest Higgs boson mass $M_{H}$ as a function of $\cos^{2}(2\beta)$.
The bottom two curves correspond to MSSM upper bounds with no threshold corrections, for $M_{top}$ = 172 GeV and 179 GeV, respectively.
 The two upper curves correspond to MSSM upper bounds with maximum threshold corrections, for $M_{top}$ = 172 GeV and 179 GeV, respectively.
The two horizontal lines are the $\cos^{2}(2\beta)$-independent SM3 vacuum stability bounds.
The lower horizontal line corresponds to $M_{top}$ = 172 GeV, while the other horizontal line was computed with $M_{top}$ = 179 GeV.
Figure 2: Same as Figure 1, but now the MSSM bounds correspond to the minimal threshold corrections consistent with the experimental lower limit on $M_{H}$
\end{quote}
\normalsize
In Fig. (1) we present our numerical two-loop results for the lightest Higgs boson mass bounds in the SM and the MSSM3 as a function of the supersymmetric parameter $\cos^{2}(2\beta)$.
The bottom two curves correspond to the MSSM3 upper bound for the two cases $M_{top} = 172$ GeV and the slightly greater upper bound that results when $M_{top}$ = 179 GeV and with no threshold corrections.
When the case of maximum threshold corrections is considered, these two curves are translated upwards by $\sim$ 55 GeV - 60 GeV, illustrating the strong dependence of the upper bound on the precise value of the threshold corrections.
Yet even with such a dramatic increase in the upper bounds with increasing threshold corrections, we observe that the SM lower bound exceeds the MSSM upper bound for $M_{top} = 172$ GeV and $ 0 < \cos^{2}(2\beta) < .2$ for all values of the threshold correction contribution.
Similarly, for $M_{top} = 179$ GeV, the troublesome situation is only exacerbated, as the SM lower bound exceeds the MSSM upper bound for $0 < \cos^{2}(2\beta) < .38$ independent of the threshold corrections.

In Fig.(2) we present the problem more clearly.
Taking into account the present experimental lower limit on $M_{H}$ of $\sim$ 90 GeV at 95$\%$ CL, we find the value of the threshold correction that gives a smallest upper bound consistent with the experimental lower limit.
Clearly, for this phenomenologically determined lower limit of the threshold contributions, there is a large area in $M_{H} \times \cos^{2}(2\beta)$ space that is inconsistent with both the SM and the MSSM.
For $M_{top} = 172$ GeV, the region 92 GeV $ < M_{H} < $ 134 GeV invalidates both theories independent of $\cos^{2}(2\beta)$, while for $M_{top} = 179$ GeV, the range of mutual invalidiation is 92 GeV $ < M_{H} < $ 150 GeV.
\section{Fourth Generation}
To resolve the above conundrum, one would like to either raise the MSSM upper bounds, lower the SM lower bounds, or both.
Upon adding a fourth generation, the SM4 lower bounds exceed the SM3 lower bounds and are an increasing function of the fourth generation masses.
If a Higgs is detected in the region of mutual invalidation of both the SM and the MSSM, consideration of SM4 vacuum stability lower bounds only exacerbates the problem.
It is readily apparent that the way out of the area of inconsistency is to consider the MSSM4 and see if the additional matter of the MSSM4 results in MSSM4 upper bounds that exceed the SM3 lower bounds.

We now discuss restrictions on the possible fourth generation fermion masses \cite{2,14,15,16}.
The close agreement betweeen the direct measurements of the top quark at the Tevatron and its indirect determination from the global fits of precision electroweak data including radiative corrections within the framework of the SM imply that there is no significant violation of the isospin symmetry for the extra generation.
Thus the masses of the fourth generation isopartners must be very close to degenerate \cite{15}; i.e.
\begin{displaymath}
\frac{\|M_{T}^{2}-M_{B}^{2}\|}{M_{Z}^{2}} \lesssim 1, \frac{\|M_{E}^{2}-M_{N}^{2}\|}{M_{Z}^{2}} \lesssim 1
\end{displaymath}
Recently, the limit on the masses of the extra neutral and charged lepton masses, $M_{N}$ and $M_{E}$, has been improved by LEP1.5 to $M_{N} > 59$ GeV and $M_{E} > 62$ GeV.
Also, CDF has yielded a lower bound on $M_{B}$ of $\sim$ 140 GeV.

In our previous work, we considered a completely degenerate fourth generation of fermions with mass $m_{4}$.
We derived an upper bound on $m_{4}$ in the MSSM4 by demanding pertubative validity of all the couplings out to the GUT scale \cite{17}.
This constraint led to an upper bound on $m_{4}$ of $\sim$ 110 GeV.
The above experimental lower limit on $M_{B}$ naturally forces us to now a consider a fourth generation where degeneracy only holds among the isodoublets seperately.
We therefore consider a fourth generation with masses $M_{L}$ and $M_{Q}$.

In Fig.(3), we present the SM lower bound, the MSSM4 upper bound with the fourth generation masses at their experimental lower limits and with fourth generation masses large enough to remove the problem area for all values of $\cos^{2}(2\beta)$.
The MSSM bounds were calculated with no threshold corrections, and $M_{top}$ is fixed at 172 GeV.
Fig.(4) shows the same information for $M_{top}$ = 179 GeV.
The MSSM4 upper bounds are much more sensitive to $M_{Q}$ than they are to $M_{L}$.
This qualitative behaviour is readily understood from inspection of the equation for $m_{\phi}^{2}$.
For this reason, it is necessary to increase $M_{Q}$ appropriately in order to generate a MSSM4 upper bound that is greater than the SM lower bound for all values of $\cos^{2}(2\beta)$.
In fact, keeping $M_{Q}$ at 146 GeV and allowing $M_{L}$ to be 110 GeV does not resolve the problem.
But increasing both $M_{Q}$ and $M_{L}$ as indicated in the figures does remove the problem.
Because all of the bounds increase as $M_{L}$ and $M_{Q}$ increase, and because the upper bounds on $m_{4}$ from the previous work are saturated when the masses of the fourth generation reach some critical values from below, we can conclude that $M_{L}$ must still be $< 110$ GeV.
This conclusion follows because it is $h_{N}$ that violates pertubative validity, so in the non-degenerate case, it is $M_{L}$ that must still respect this upper bound if gauge coupling unification is still to be achieved in the MSSM4.

\hglue -2cm
\psfig{figure=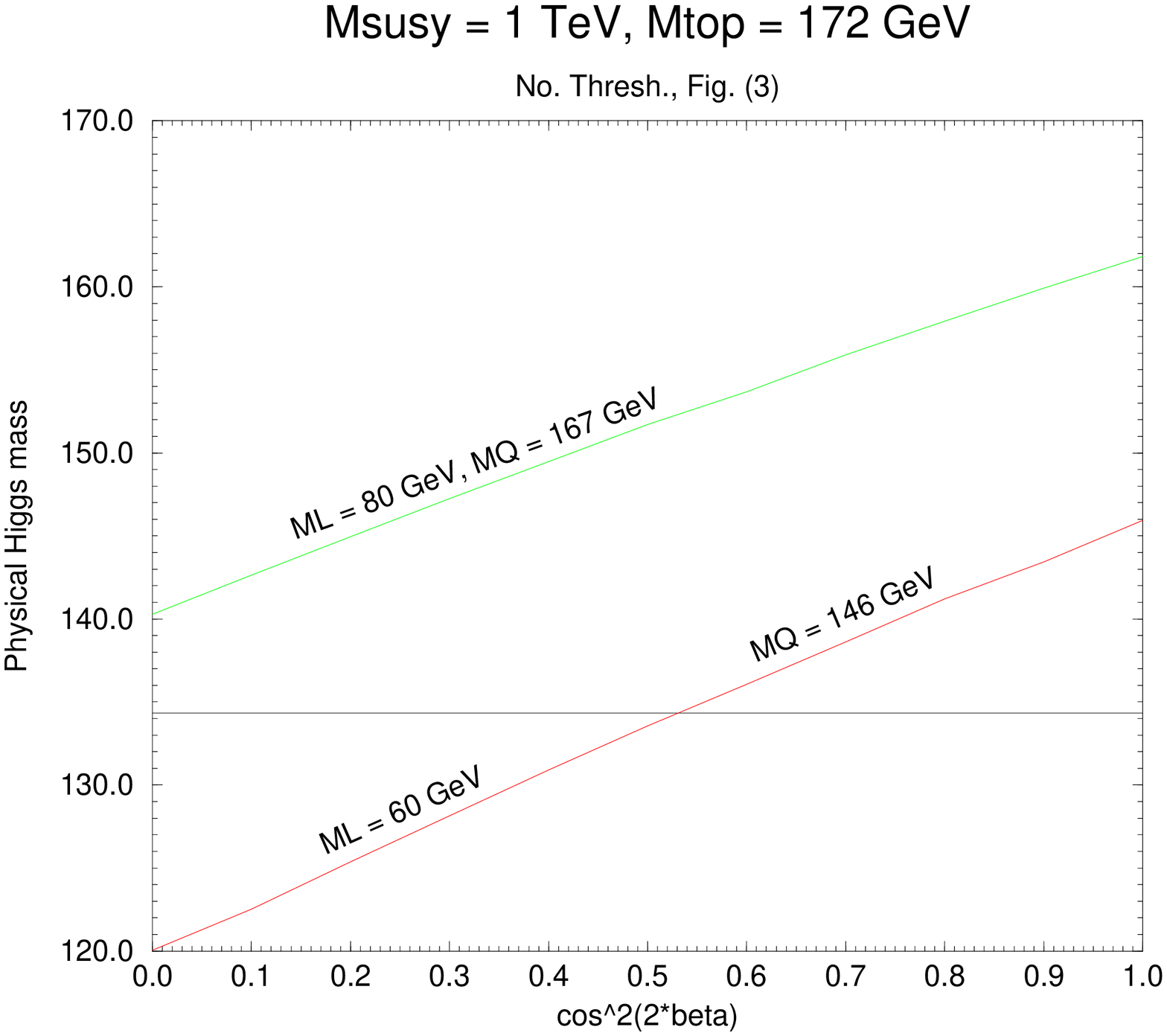,height=8cm,angle=-90}\hglue 1cm
\psfig{figure=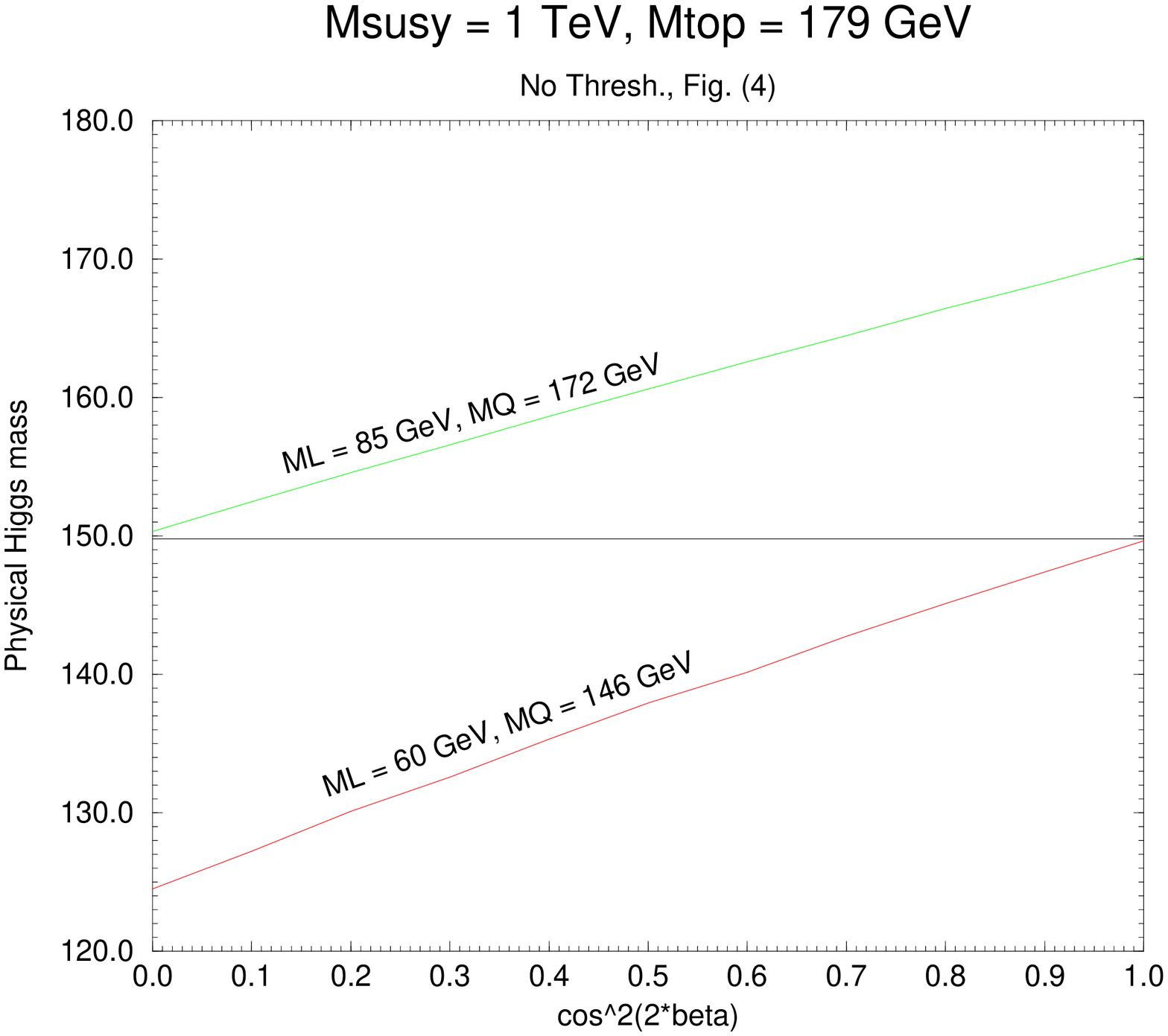,height=8cm,angle=-90}
\hglue 3.5cm (3)\hglue 10cm (4)\\
\begin{quote}
\scriptsize Figure 3: Plots of the physical Higgs boson mass as a function of $\cos^{2}(2\beta)$. The $\cos^{2}(2\beta)$-independent flat line is the MSSM3 vacuum stability lower bound for $M_{top}$ = 172 GeV. The lower curve is the MSSM4 upper bound for the same value of $M_{top}$, no threshold corrections and the indicated values for $M_{L}$ and $M_{Q}$. Similarly for the upper curve. Figure 4: Same as Figure 3, but with $M_{top}$ = 179 GeV.
\end{quote}
\normalsize
\section{CONCLUDING REMARKS}

In conclusion, we have studied the upper bounds on the lightest Higgs boson mass $M_{H}$ in the MSSM with four generations by solving the two-loop RGE's and using the one-loop EP.
We find that if the Higgs boson is discovered with a mass $M_{H} < $134 GeV (150 GeV) for $M_{top} =$ 172 GeV (179 GeV), then there is a demand for the introduction of new physics.
This mass range for $M_{H}$ will be explored shortly and thus an explanation of what new physics could be consistent with such a Higgs mass measurement is desirable.
We propose that such a measurement could be taken as indirect evidence for a fourth generation of fermions.
Considering a fourth generation where degeneracy only holds within the isodoublets individually, we find that a measurement of $M_{H}$ in the above range is consistent with the MSSM4 upper bounds on $M_{H}$.
In addition, the possibility of gauge coupling unification remains intact for 60 GeV $< M_{L} <$ 110 GeV and $M_{Q} \stackrel{>}{_{\sim}}$ 170 GeV.
Therefore, if $M_{H}$ is measured to be below the SM3 lower bound, we suggest a search for fourth generation fermions with 60 GeV $< M_{L} <$ 110 GeV and $M_{Q} \stackrel{>}{_{\sim}}$ 170 GeV. 

\section{ACKNOWLEDGEMENTS}

We wish to thank M. Machacek and M. Vaughn for helpful discussions concerning the RGE used in this investigation.
Support for this work was provided in part by U.S. Dept. of Energy Contract DE-FG-02-91ER40688-Task A.


\newpage


\begin{thebibliography}{99}

\bibitem{1} D. Dooling, K. Kang, and S.K. Kang, IJMPA (in press), preprint hep-ph/9710258.
\bibitem{2} S.K. Kang and G.T. Park, Mod. Phys. Lett. A 12 (1997) 553.
\bibitem{3} S.K. Kang, Phys. Rev. D 54 (1996) 7077.
\bibitem{4} J. Kodaira, Y. Yasui, and K. Sasaki, Phys. Rev. D 50 (1994) 50.
\bibitem{5} M. Sher, Phys. Rep. 179 (1989) 273, and references therin.
\bibitem{6} J.A. Casas, J.R. Espinosa, M. Quiros, A. Riotto, Nucl. Phys. B 436 (1995) 3; B 439 (1995) 466 (E).
\bibitem{7} J.A. Casas, J.R. Espinosa, and M.Quiros, Phys. Lett. B 342 (1995) 171; B 382 (1996) 374.
\bibitem{8} Y. Okada, M. Yamaguchi, and T. Yanagida, Phys. Lett B 262 (1991) 54; Prog. Theor. Phys. 85 (1991) 1; H.E. Haber and R. Hempfling, Phys. Rev. Lett. 66 (1991) 1815; Phys. Rev. D 48 (1993) 4280; J. Ellis, G. Ridolfi, and F. Zwirner, Phys. Lett. B 257 (1991) 83; R. Barbieri, M. Frigeni, and F. Caravaglios, Phys. Lett. B 258 (1991) 167.
\bibitem{9} G. Altarelli and G. Isidori, Phys. Lett. B 337 (1994) 141.
\bibitem{10} C. Ford, D.R.T. Jones, P.W. Stephenson, and M.B. Einhorn, Nucl. Phys. B 395 (1993) 17.
\bibitem{11} M. Machacek and M. Vaughn, Nucl. Phys. B 222 (1983) 83; B 236 (1984) 221; B 249 (1985) 70.
\bibitem{12} M. Lindner, Z. Phys. C 31 (1986) 295.
\bibitem{13} J. R. Espinosa and M. Quiros, Phys. Lett. B353, 257 (1995).
\bibitem{14} H.B. Nielsen, A.V. Novikov, and M.S. Vysotsky, Phys. Lett. B 374 (1996) 127.
\bibitem{15} V. Novikov, preprint hep-ph/9606318 (June 1996); LEP1.5 Collaboration, J. Nachtman, in Electroweak Interactions and Unified Theories, Proceedings of the 31st rencontres de Moriond, Les Arcs, France (1996), preprint hep-ex/960615.
\bibitem{16} K.S. Babu and E. Ma, Z. Phys. C 29 (1985) 45.
\bibitem{17} J.F. Gunion, D.W. McKay, and H. Pois, Phys. Lett. B 334 (1994) 339; J.F. Gunion, D.W. McKay, and H. Pois, Phys. Rev. D 53 (1996) 53.
\end{thebibliography}
\end{document}